\documentclass[sigconf]{acmart}

\AtBeginDocument{%
  }

\setcopyright{acmcopyright}
\copyrightyear{2024}
\acmYear{2024}
\acmDOI{XXXXXXX.XXXXXXX}

\begin{document}

\title{Investigating the use of Snowballing on Gray Literature Reviews}
\subtitle{A Study on Stack Exchange}


\author{Felipe Gomes}
\email{felipe.gustavo@ufba.br}
\authornotemark[1]
\affiliation{%
  \institution{Federal University of Bahia and Federal Institute of Bahia}
  \country{Brazil}
}

\author{Thiago Mendes}
\email{thiagosouto@ifba.edu.br}
\affiliation{%
  \institution{Federal Institute of Bahia}
  \country{Brazil}
}

\author{Sávio Freire}
\email{savio.freire@ifce.edu.br}
\affiliation{%
  \institution{Federal Institute of Ceará}
  \country{Brazil}
}

\author{Rodrigo Spínola}
\email{spinolaro@vcu.edu}
\affiliation{%
  \institution{Virginia Commonwealth University}
  \country{Brazil}
}

\author{Manoel Mendonça}
\email{manoel.mendonca@ufba.br}
\affiliation{%
  \institution{Federal University of Bahia}
  \country{Brazil}
}

\renewcommand{\shortauthors}{Gomes et al.}

\begin{abstract}

\textit{\textbf{Background}}: The use of gray literature (GL) has grown in software engineering research, especially in studies that consider Questions and Answers (Q\&A) sites, since software development professionals widely use them. Though snowballing (SB) techniques are standard in systematic literature reviews, little is known about how to apply them to gray literature reviews. \textit{\textbf{Aims}}: This paper investigates how to use SB approaches on Q\&A sites during gray literature reviews to identify new valid discussions for analysis. \textit{\textbf{Method}}: In previous studies, we compiled and analyzed a set of Stack Exchange Project Management (SEPM) discussions related to software engineering technical debt (TD). Those studies used a data set consisting of 108 valid discussions extracted from SEPM. Based on this start data set, we perform forward and backward SB using two different approaches: link-based and similarity-based SB. We then compare the precision and recall of those two SB approaches against the search-based approach of the original study. \textit{\textbf{Results}}: In just one snowballing iteration, the approaches yielded 291 new discussions for analysis, 130 of which were considered valid for our study. That is an increase of about 120\% over the original data set (recall). The SB process also yielded a similar rate of valid discussion retrieval when compared to the search-based approach (precision). \textit{\textbf{Conclusion}}: This paper provides guidelines on how to apply two SB approaches to find new valid discussions for review. To our knowledge, this is the first study that analyzes the use of SB on Q\&A websites. By applying SB, it was possible to identify new discussions, significantly increasing the relevant data set for a gray literature review. 

\end{abstract}


\maketitle

\section{Introduction} \label{sec:introduction}


Grey literature (GL) encompasses publications not subject to formal quality control mechanisms, such as peer-review, before its release~\cite{petticrew2008systematic}. Different research areas, such as medicine and nutrition, have been using GL as a valuable source of professional knowledge~\cite{kamei2021grey}. GL often describes user-generated web content, like tweets, blogs, and posts on question and answer (Q\&A) websites. Recently, the interest in GL has grown in the context of Software Engineering (SWE) research, mainly because it is widely used by SWE practitioners, especially the Q\&A platforms.  

In SWE, GL is often used to discuss various software engineering topics, with software professionals sharing their knowledge or asking for advice or solutions on software development issues. SWE researchers have noticed that GL is a rich source of SWE knowledge, used alone or combined with findings published in the formal white literature (WL)~\cite{kamei2021grey}.

Q\&A platforms, such as the \href{https://stackexchange.com/}{Stack Exchange} (SE), are part of daily activities of modern software development, and these platforms have raised the interest of software researchers \cite{garousi2019guidelines}. They are a rich source of information because countless issues are discussed in them, bringing to light the point of view of practitioners on possible solutions for those issues~\cite{vasilescu2014social}. Moreover, those platforms usually count on a self-moderating system for the discussions and users, making the available information self-regulated and more reliable for practitioners and, by proxy, researchers.

SE is composed of different sites, each one focused on a specific topic, such as software development (\href{https://stackoverflow.com/}{Stack Overflow}) or project management (\href{https://pm.stackexchange.com/}{Stack Exchange Project Management - SEPM}). Researchers have analyzed discussions from SE regarding topics such as code smells~\cite{tahir2020large}, software development trends~\cite{barua2014developers}, psychological safety~\cite{Santana2024}, software requirements~\cite{freire2023}, soft skills~\cite{montandon2021skills}, and technical debt (TD)~\cite{digkas2019reusing, Gama2020, blind-1, blind-2, blind-3}. 

In all those studies, GL reviews have relied on keyword searches to obtain the information they want extracted for analysis. In contrast, systematic literature review (SLR) studies, which are very common in SWE, combine database searches and snowballing (SB) as the primary strategy for conducting the searches in these types of studies~\cite{badampudi2015experiences}.  

In SLR, researchers search different databases using predefined search strings to identify new papers~\cite{jalali2012systematic} and use the SB method to expand the pool of relevant studies~\cite{wohlin2014guidelines}. SB recursively traces the references list and citations of initially retrieved papers, also called start set, to identify additional sources that might have been missed~\cite{wohlin2014guidelines}. These strategies have been the target of research that seeks to define guidelines for SB~\cite{badampudi2015experiences, wohlin2014guidelines}, compare SB to database searches~\cite{jalali2012systematic}, and propose a hybrid approach combining the two approaches~\cite{wohlin2022successful}.

Despite the use of SB techniques being common in SLR studies and the interest of the SWE community in GL, to the best of our knowledge, there are no studies analyzing how SB approaches can be translated to Q\&A websites, and analyzing the precision and recall of these approaches. This paper investigates how SB techniques can be applied to SE websites to discover new and valid discussions for GL reviews.

We have previously analyzed Stack Exchange's SEPM, which encompasses discussions of practitioners interested in project management, to investigate how project managers discusses and experiences TD. This previous work analyzed 108 TD-related discussions, accounting for 547 posts and 882 text comments. This previous work, which did not use SB, is reported in three papers~\cite{blind-1, blind-2, blind-3}.

 
Based on this data set, we performed SB, resulting in a data set with 291 new discussions, composed of 1985 posts and 2532 comments, from SEPM.  In order to evaluate the precision and recall of the SB, we analyzed the 291 discussions quantitatively and qualitatively using the same guidelines from our previous studies~\cite{blind-1, blind-2, blind-3}. Out of 291, we found 130 new valid discussions using backward and forward SB, using two different approaches: link-based SB and similarity-based SB. The amount of new discussions represents an increase of about 120\% about the initial data set. Surprisingly, the similarity-based SB yielded good results compared to link-based SB and the original search-based approach.

It is important to point out that the results reported here will not focus on the TD research outcome. Instead, this paper focuses on the SB approach we used, the validity of the discussions we found, and the insights we gained on this process.

The remainder of the paper is organized as follows. Section~\ref{sec:related-work} presents related work. Section~\ref{sec:start-set-and-analysis-process} describes how we extracted and analyzed the start set from the original studies. Section~\ref{sec:research-method} presents how we adapted the traditional snowballing approach to SWE Q\&A websites. Section~\ref{sec:result-and-discussion} presents the results and discussion of the quantitative and qualitative analyses. Section~\ref{sec:threats-to-validity} discusses the threats to the study's validity. And lastly, Section~\ref{sec:conclusion} presents our final remarks.

\section{Related Work} \label{sec:related-work}


SWE research widely employs SLR studies. They are commonly used to understand or organize the available knowledge in a research area. When conducting SLR studies, researchers usually use SB in addition to database search to find more relevant papers to their data set. Wohlin \textit{et al.}~\cite{wohlin2014guidelines} have proposed guidelines to conduct SLR studies using SB. These guidelines were based on the experiences gained over multiple SLR studies. They concluded that using SB, as a first search strategy, may very well be a good alternative to the use of database searches. Wohlin \textit{et al.}~\cite{wohlin2022successful} also proposed and evaluated a hybrid strategy, combining database search and SB. They concluded that the hybrid strategy is very effective in finding relevant studies. Deepika \textit{et al.}~\cite{badampudi2015experiences} evaluated the precision and reliability of SB as a search strategy in literature studies. By comparing the findings of SB approaches and database search, they concluded that the precision of SB is comparable to that of a database search. Furthermore, they also conclude that SB can be more reliable than a database search. However, the reliability is highly dependent on creating a suitable start data set.

GL reviews have gained increased interest in SWE research in recent years. In 2021, Kamei \textit{et al.}~\cite{kamei2021grey} published a tertiary study summarizing the last ten years of SWE research that uses GL, showing that GL has been essential for bringing practical new perspectives that are scarce in traditional literature. They drew the current use landscape and raised awareness of challenges related to GL reviews.

With the increased interest in GL, SWE researchers expanded on how GL should be used in SWE literature reviews. Garousi \textit{et al.}~\cite{garousi2019guidelines} proposed multivocal literature review (MLR) guidelines to incorporate GL alongside traditional literature. The authors compared the results of two investigations: one included GL, while the other did not. Their findings highlighted the importance of using GL to cover technical research questions. Raulamo-Jurvanen \textit{et al.}~\cite{raulamo2017choosing} conducted a grey literature review (GLR) aiming to understand how software practitioners choose the right test automation tool. Their findings are mainly derived from practitioners' experiences and opinions. To improve the credibility of their findings, they employed some criteria to assess the evidence of the GL, including, for instance, the number of readers, the number of comments, or the number of hits on Google.

Researchers have also been using (Q\&A) platforms to empirically understand how practitioners manage TD, a research topic that is the focus of our original GL-based research~\cite{blind-1, blind-2, blind-3}. Digkas \textit{et al.}~\cite{digkas2019reusing} conducted an empirical study on the relation between the existence of reusing code retrieved from Stack Overflow on the TD of the target system. The results provide insights into the potential impact of small-scale code reuse on TD and highlight the benefits of assessing code quality before committing changes to a repository. In another work, Gama \textit{et al.}~\cite{Gama2020} investigated the point of view of practitioners on Stack Overflow on how developers commonly identify TD items in their projects. They reported that developers commonly discuss TD identification, revealing 29 low-level indicators for recognizing code, infrastructure, architecture, and test debt items.

Despite the SWE research community's interest in GL, to our knowledge, no previous work has investigated how SB can be applied to Q\&A platforms to discover new and valid discussions for GL reviews. This knowledge gap is precisely what this article addresses.
 
As mentioned before, we have recently performed three studies investigating, from the point of view of project management practitioners, how they commonly discuss, experience, and manage TD on SEPM \cite{blind-1, blind-2, blind-3}. We identified 74 indicators project managers use to recognize debt items and 126 TD management practices. The types of debt most discussed at SEPM are process and people debt, which contrasts with previous TD investigations that focused mainly on software develpers \cite{Rios2018}, including a GL TD study that analyzed StackOverflow \cite{Gama2020}. There, the software developers talk mostly about code and design debt.

The inclusion and analysis criteria used in our SB study are the same as those used in the abovementioned studies~\cite{blind-1, blind-2, blind-3}. We will discuss them in the following Section.

\section{Start set and Analysis Process} \label{sec:start-set-and-analysis-process}


The basis of every SB procedure is the start set~\cite{wohlin2022successful}. We use the data set we built in our previous studies on TD~\cite{blind-1, blind-2, blind-3} as our start set. To prepare this data set, we used \href{https://archive.org/details/stackexchange}{SE data dump} (version Sep 07, 2021), which is the raw data dump of all SE websites. From this repository, one can access data from any SE website, including posts (both questions and answers), comments, and metadata. In the next sections, we present the components of a discussion in SE websites, in addition to the processes of collection and analysis of these discussions.

\subsection{Discussion Structure}

\begin{figure*}[htp]
\centerline{\includegraphics[width=\linewidth]{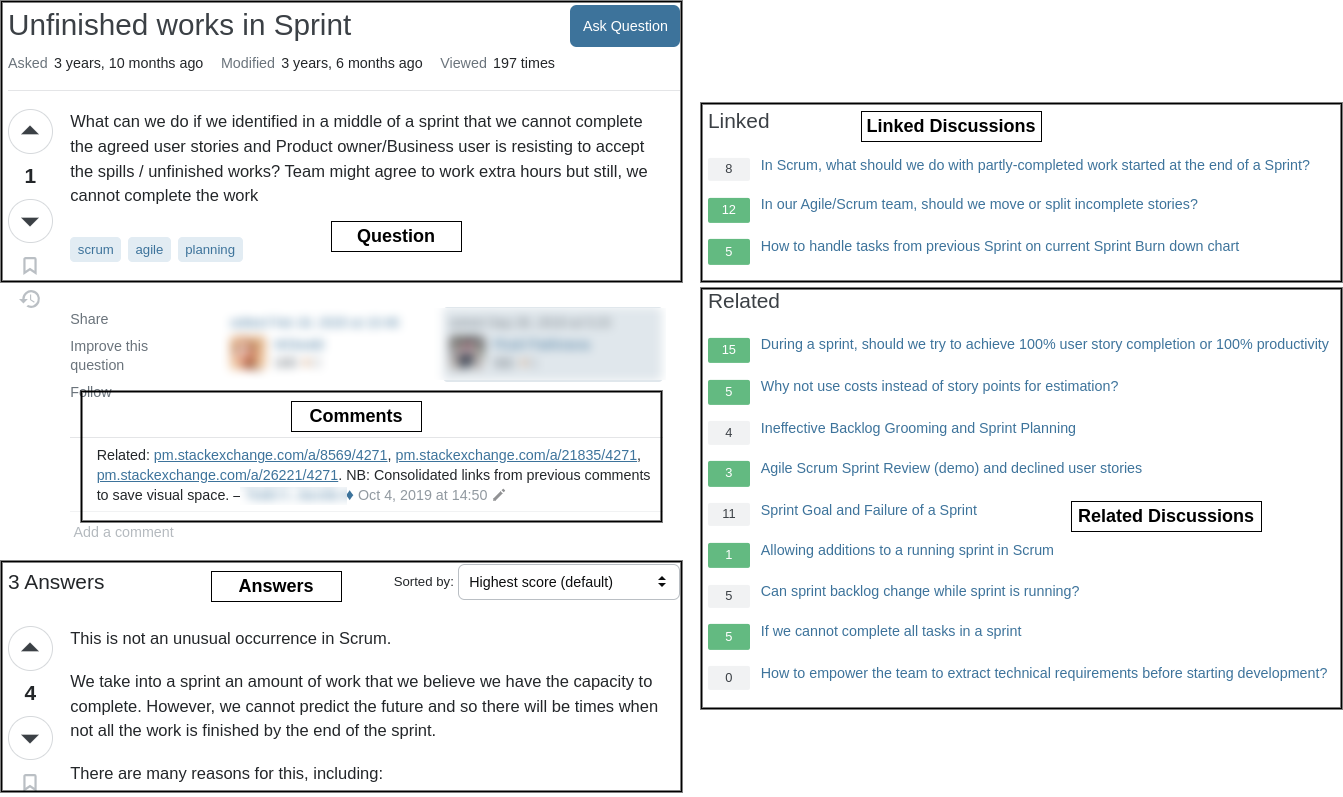}}
\caption{Example of a discussion.}
\label{fig-1}
\end{figure*}

The discussions in SE websites are composed of three main components: the question around which the discussion is centered, the answers to the question, and the comments on both questions and answers. Comments are presented below the post they are related to. Thus, they are part of the discussions. Figure~\ref{fig-1} shows, on the left, an example of a discussion presenting these components. 

Besides these main components, the discussions also may have two types of links to other discussions: linked and related. In Figure~\ref{fig-1}, we can see these links on the right.

The linked list gathers any links to a specific post the community provides via comments, answers, or questions. In other words, when a user puts a link in the discussion pointing to another discussion in the same forum, a linked link is created. The link is still listed even if the users remove a post or comment. In Figure~\ref{fig-1}, we have an example of a link to another discussion in the comment on the question. The links are akin to references that the users purposefully placed during the SE discussions.

The related links were introduced in the \textit{`StackExchange 2.0'}. This type of link points to discussions classified as `related'. To carry out this classification, the SE relies on the `More like this' query provided by Elasticsearch, which is a search engine and analytic tool that provides features like \href{https://www.elastic.co/what-is/elasticsearch}{full-text search}. The `More like this' query is based on the Term Frequency - Inverse Document Frequency (TF-IDF) algorithm. The TF-IDF is a numerical statistic that reflects how important a word is to a document in the collection or corpus~\cite{salton1988term}.


The related links are generated automatically by the SE platform. It is performed as an adapted TF-IDF using the following criteria: full-text match to tags (+10 weight), full-text match to title (+5 weight), and full-text match to body (+1 weight)~\cite{SELinks}. Unfortunately, other details about the related link list generation are closed and subject to change over time. Therefore, we are unable to have a complete understanding of how the algorithm used by SE works.

\subsection{Data Collection}

To build the start set~\cite{wohlin2022successful}, we searched the SEPM content for the strings ``debt'' or ``shortcut'' in the questions (title, body, and tags), answers’ body, and comments’ body. The term ``debt'' catches the different forms of reference to TD, for example, ``tech debt'' and ``code debt''. By performing pilot studies, we also detected that the term ``shortcut'' was used to refer to TD in the discussions. The term ``shortcut'' catches other references to sub-optimal solutions.

Our previous works~\cite{blind-1, blind-2, blind-3} included the comments in the analyses because they provide helpful information relevant to identifying TD indicators and management practices, such as scenario context. Examining comments is particularly relevant when considering discussions on SE websites because comments go beyond the discussion, clarifying and enriching the content conveyed through questions and answers \cite{sengupta2020learning}.

We choose such general terms because of the free nature of discussion forums. The forum user can refer to TD in several ways, and unlike formal literature, there is no standard. Some users refer to TD as ``tech debt'', ``development shortcut'', ``delivery shortcut'', or just ``debt''. To decide which strings to use, we tested some strings like ``technical debt'', ``code debt'' and other terms used in the literature, but the number of returned discussions was very low. Therefore, after the tests, we decided to use the terms ``debt'' and ``shortcut'' because they encompassed the discussions found using the other terms and returned a higher number of discussions. 

\begin{figure}[htp]
\centerline{\includegraphics[width=\linewidth]{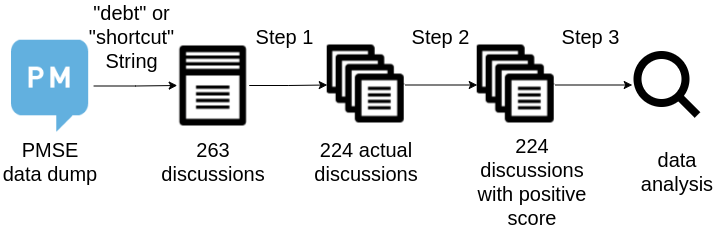}}
\caption{Data extraction and filtering process.}
\label{fig-data-extraction}
\end{figure}

The string-matching phase yielded a total of 263 discussions. Next, we filtered the data as shown in Figure \ref{fig-data-extraction}, using the following criteria:

\begin{itemize}
\item \textbf{Step 1:} \textit{Eliminate incomplete discussions from the data set}. We considered a discussion complete when a question was followed by one or more answers, where there was at least one answer whose author differed from the question's author. After applying this criterion, 224 discussions remained out of the initial 263.
\item \textbf{Step 2:} \textit{Eliminate untrustworthy discussions}. Other studies have found that data from Q\&A forums can be affected by noise~\cite{ahasanuzzaman2016mining, kavaler2013using}, requiring mitigating it using different proxies~\cite{Gama2020}. We decided to use the discussion score as a filtering proxy. A post score is a SE popularity metric in which users, other than the post author, can give an up-vote to the post if they find it useful or a down-vote if they find it not useful. A discussion score is the difference between up-votes and down-votes of all its posts. We decided to filter out discussions with negative scores, since overall they are not considered useful. While we did not discard any discussion in this step, we considered it important since can help to ensure the quality of the discussions. 
\item \textbf{Step 3:} \textit{Qualitative data analysis.} We conducted a qualitative data analysis of the data set. Each of the 224 discussions went through a qualitative analysis process.
\end{itemize}

\subsection{Data Analysis}

\begin{figure}[htp]
\centerline{\includegraphics[width=\linewidth]{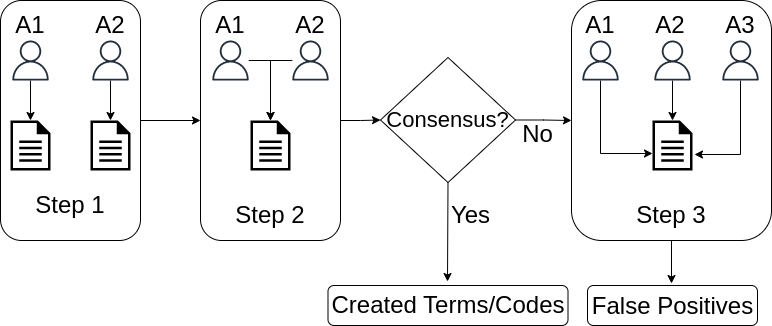}}
\caption{Discussions analysis procedure.}
\label{fig-analysis-process}
\end{figure}

We performed manual qualitative analysis~\cite{seaman1999qualitative} over the data set. The analysis was composed of three steps shown in Figure~\ref{fig-analysis-process}. In~\textbf{step 1}, the first two authors of this paper analyzed each discussion independently. For each discussion, individually, they filled in the set of questions presented in Table \ref{tab-checklist}, reporting the types of debt, TD indicators, technical debt management (TDM) practices, agile roles, artifacts, and events affected by the TD presented in each discussion.

During \textbf{step 1}, when information was found, it was recorded according to the perception of each researcher, without any previously established codes. The researchers were encouraged to freely record information regarding TD, adding as much information as they deemed necessary to answer the questions in Table~\ref{tab-checklist}. We decided to use this approach because the analysis required a fair amount of understanding and interpretation, and we wanted to keep the initial data acquisition as simple as possible. This approach was considered most applicable given the lack of previous analysis of the discussions and the nature of the discussions themselves. 

The exception to the above-described procedure was using predefined codes for \textit{TD types}. To identify the type of debt addressed in the discussions, we employed the definitions reported by Rios \textit{et al.}~\cite{Rios2018}. As an example, let us consider the following discussion excerpt: \textit{``I suspect the team is either interviewing or ramping up their skills to prepare for their next job. How do I motivate them to restore their velocity?''} This discussion occurs in the context of problems that might happen due to overloaded individuals, characterizing a discussion regarding people debt, as defined by Rios \textit{et al.}~\cite{Rios2018}.

\begin{table}[htp]
\caption{Questions to gather TD information in SEPM discussions.}
\begin{center}
\begin{tabular}{l}
\hline
\textbf{Questions (Q)}\\
\hline
\textbf{Q1.} What types of TD are discussed?\\
\hline
\textbf{Q2.} What are the discussed indicators of TD?\\
\hline
\textbf{Q3.} What are the activities and strategies that were  discussed\\to support TD management and identification?\\
\hline
\end{tabular}
\label{tab-checklist}
\end{center}
\end{table}

In this information-gathering effort, \textbf{Q1} captures the type of debt discussed by SEPM's users, according to the concepts presented by Rios \textit{et al.}~\cite{Rios2018}. \textbf{Q2} captures the TD indicator reported by SEPM users to recognize TD items. Lastly, \textbf{Q3} catches TDM practices proposed by SEPM users to address the identified TD items.  

While going through the information-gathering questions (\textbf{Step 1}), we also looked for false positives, taking into consideration the following rules:

\begin{itemize}
\item \textbf{Rule 1:} \textit{The discussion must be related to TD.} Discussions that, in spite of having the terms \textit{``debt''} or \textit{``shortcut''}, did not discuss TD were marked as false positives.
\item \textbf{Rule 2:} \textit{The discussed situation must be real.} This work intends to map real problems faced by practitioners, so questions asking for advice without bringing any actual situation from the present or past were flagged as false positives.
\item \textbf{Rule 3:} \textit{The TD indicators must come from the question's author.} Since the question's author is the one with actual knowledge about the situation, we only considered the author's words concerning TD indicators. TD indicators inferred by other users were considered only if sustained by the question's author in a comment or a post in the same discussion.
\item \textbf{Rule 4:} \textit{The recommended TDM practices needed to be backed by the community}. We intend to map practices viewed as good by the community, thus, we only considered practices recommended in answers or comments that were backed by other users than the recommendation's author. Therefore, practices needed to be accepted by at least one other user quantitatively (giving a positive score) or qualitatively (giving a favorable opinion) to be considered.
\end{itemize}

The combination of the rules defined above and the questions presented in Table \ref{tab-checklist} allowed us to verify whether a discussion was within the scope of information gathering or was a false positive. After removing all false positives, our final data set was reduced to 108 discussions containing 547 posts and 882 comments. Of those, 81 discussions were related to agile software development (436 posts and 691 comments), and 27 were unrelated to agile software development (111 posts and 191 comments).

We used \textbf{step 2} to examine the responses recorded individually in the first step and reach a consensus between the researchers. Together, the two researchers checked the information extracted from the discussions and in the case of analysis agreement, they also performed the coding process. Discussions that yielded divergent information were subjected to a second examination in \textbf{step 3}, this time with a third researcher. A majority vote among the researchers would define which information should remain for the following analysis step. This process was repeated until all responses were consolidated.

For each discussion, we created terms to represent and group the TD indicators that were addressed in various ways. For example, the following TD indicator \textit{``I've been appointed as a Scrum Master to a new team and senior management (CTO) expectation is that our team should deliver more than the SP capacity we can currently realistically deliver.''} was coded to \textit{``The agile practices are not respected.''}. Whenever we found TD indicators related to more than one TD type, we considered them associated with all TD types. For example, the indicator \textit{``The agile practices are not respected.''} is at the same time related to process and people debt. The relation to process debt is due to the process that lets the problem occur, and the relation to people debt is due to the practitioners deciding not to follow the agile practices.

Once again, disagreements in the coding were resolved with the help of the third researcher. This process was performed until no new codes were identified (point of saturation). Then, this encoded data set was added to the results. Besides that, we also performed a grouping of the TDM practices following the same approach, by a consensus of two researchers and using a third researcher to help with disagreements.

\begin{figure}[hbp]
\centerline{\includegraphics[width=\linewidth]{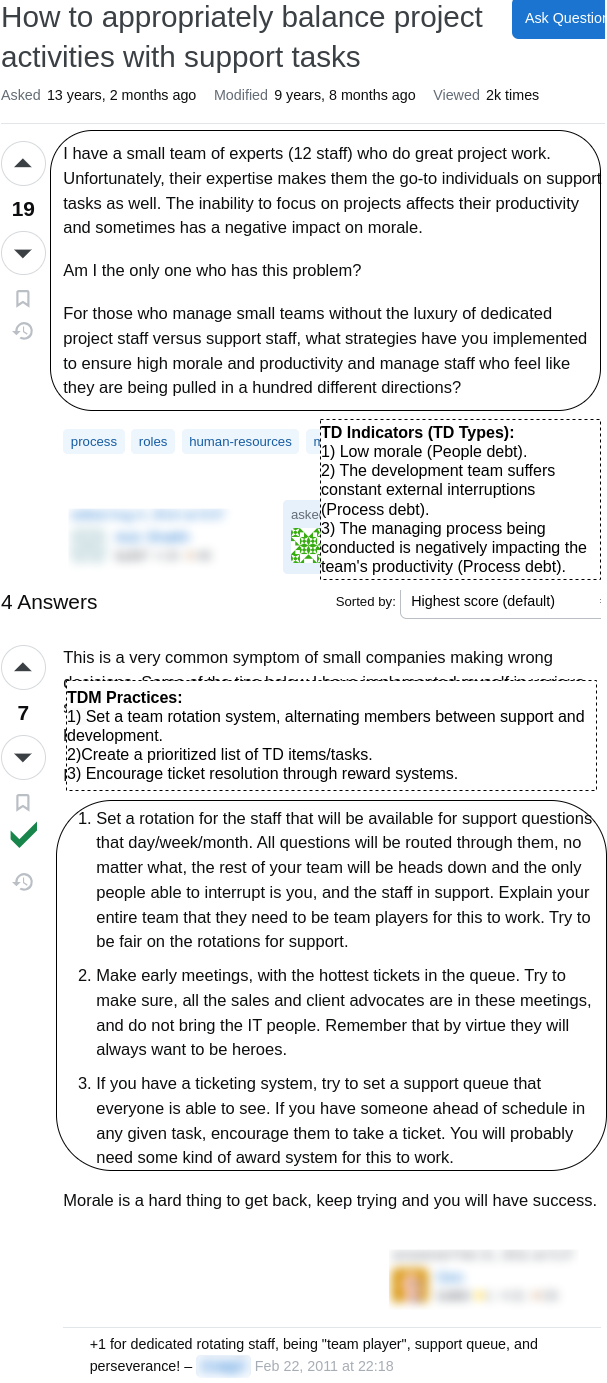}}
\caption{Example of a valid discussion.}
\label{fig-example-analysis-1}
\end{figure}

An example of an analysis of a discussion of interest is shown in Figure~\ref{fig-example-analysis-1}. Based on the question's body, we marked the discussion as process and people debt. Corroborating with the marked TD types, the question's body gives more context to the problem, helping us to identify three TD indicators, \textit{``Low morale.''}, \textit{``The development team suffers constant external interruptions.''} and \textit{``The managing process being conducted is negatively impacting the team's productivity.''}. From the question's content, we can assume that what is harming the development productivity is the management decisions.

The answers in Figure \ref{fig-example-analysis-1} indicate three TDM practices that can be applied to mitigate the problem under discussion. The TDM practices revolve around changes in the management process to address the TD. The recommended TDM practices in the discussion were: \textit{``Set a team rotation system, alternating members between support and development.''}, \textit{``Create a prioritized list of TD items/tasks.''} and \textit{``Encourage ticket resolution through reward systems.''}. This discussion is an example of the ``TD-related problems and practices to mitigate these problems'' structure we were looking for during the analysis. Through it, we can trace the relationship between the TD indicators and TDM practices.

Regarding the usefulness of comments in the analysis. Figure \ref{fig-example-analysis-1} shows a comment attached to an answer. Based on both the comment provided by the question author (attached to the answer) and the answer accepted mark (the green check), we can see that the author approved the suggested solutions.

\begin{figure}[htp]
\centerline{\includegraphics[width=\linewidth]{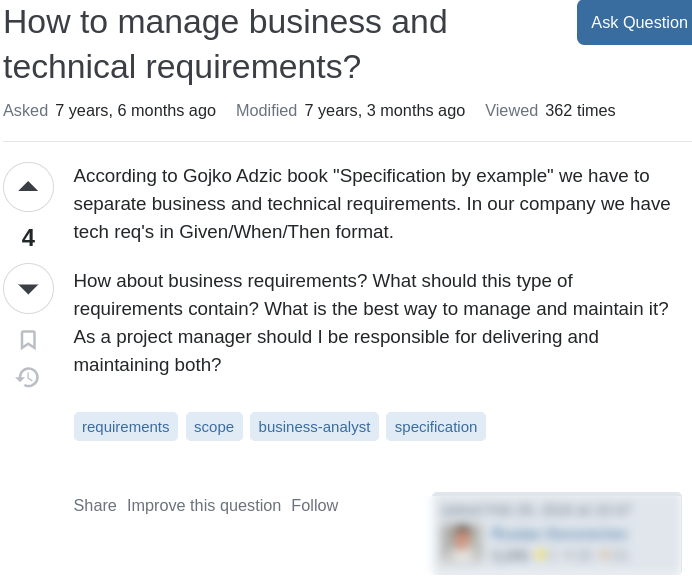}}
\caption{Example of false positive discussion.}
\label{fig-example-false-positive}
\end{figure}

In Figure~\ref{fig-example-false-positive}, we have an example of a discussion flagged as false positive. The discussion is about a theoretical doubt regarding requirements management and is not related to a real-world scenario.

It is worth mentioning that the extraction of the TD-related data was not a straightforward task. We needed to interpret the posts to perform the data extraction and coding. This required us to reread posts several times to clear any doubts about their content. We also consulted the links to other web pages, such as articles and tutorials. Take the following discussion excerpt as an example: \textit{``Here is a good article I wrote on my experience with the topic: I solved the Agile testing bottleneck problem! link\footnote{\url{https://medium.com/@salibsamer/i-solved-scrum-sprint-end-testing-bottleneck-problem-bfd6222284a1}}''}. There was also the necessity to understand the jargon used by the practitioners, such as \textit{``low hanging fruit''}\footnote{\url{https://aiko.dev/visualising-and-prioritizing-technical-debt/}}, \textit{``gold plating''}\footnote{\url{https://www.linkedin.com/pulse/gold-plating-software-engineering-project-management-chirag-pmp}}, and others, that required further reading from external sources.

After the analysis process, 108 discussions remained and these discussions composed the start set. This start set was then used in the SB processes described in the next sections.

\section{Snowballing Study Methodology} \label{sec:research-method}

This section presents the research questions along with how we adapt the snowballing approach to the SE websites.

\subsection{Research Questions}

As previously stated, this work aims to investigate how SB techniques can be applied in SE websites to discover new valid discussions. To this end, we seek answers to the following main research questions (RQs):

\begin{itemize}
    \item \textbf{RQ1:} \textit{How many new discussions does the snowballing find in relation to the original data set?} - This question seeks to identify how many new valid discussions can be discovered in SEPM through SB. This metric is relevant, given that in the WL, SB is widely used to expand the initial pool of papers.
    \item \textbf{RQ2:} \textit{How precise are snowballing techniques in finding new valid discussions?} - This question aims to identify the number of valid discussions in relation to the new discussions discovered in the SEPM through SB. As pointed out by Wohlin~\cite{wohlin2014guidelines}, one important efficiency measure for SLRs is the number of included papers in relation to the total number of candidate papers examined. Thus, investigating the precision of SB in GL is also relevant.
\end{itemize}

About to the RQs, the following contributions are made:
\begin{itemize}
    \item A method proposal on how SB can be performed in Q\&A websites and an evaluation of the effectiveness of this method. The SB conducted rendered 20\% more valid discussion than the original study, and hence it provides an added value to the original study. 
    \item Two strategies of SB are compared, linked SB and related SB. Related SB found more discussions than linked SB, but they are complementary. It is concluded that the full-fledged hybrid search strategy is the best, although it comes at the cost of more work.
\end{itemize}

The following sections explain how we adapted and applied snowballing techniques to discussion forums to answer the \textbf{RQs}.

\subsection{Snowballing Procedure}

The basic planning and motivation of a systematic literature study are independent of the search approach. Thus, the basic steps for planning a literature study presented in Keele \textit{et. al}~\cite{keele2007guidelines} are still relevant even if applying a different approach to the search. 

Snowballing is a widely used technique in the white literature. It refers to using the reference list of a paper (backward SB) or the citations to the paper (forward SB) to identify additional papers. Snowballing is conducted by first creating a start set, which is a set of articles around a topic of interest, then the two approaches can be applied: Forward Snowballing (FSB) and Backward Snowballing (BSB). Once the start set is decided, it is time to start the BSB and FSB. 

In traditional BSB, one use the reference list of the start set to identify new papers to include. On the other hand, FSB refers to identifying new papers based on those papers citing the papers within the start set. 


To map the traditional snowballing approach to discussion forums, we considered the GL discussions as papers and the links between them (related and linked links) as references from one paper to another. We adapted the traditional snowballing approach to discussion forums by considering these links akin to references or citations between articles. In GL, it is crucial to have inclusion criteria before analyzing and coding the whole discussion.

Regarding the linked links, we tried to obtain them in two ways. Firstly, we mined the links using the {SE data dump} (version April 06, 2024). The links are stored in a specific SE table, named PostLinks. It remains recorded even when the post or comment containing that link gets deleted. This table contains two types of links: linked (a post contains a link to another post) and duplicate (a post has a link to itself). The duplicate links were not considered during the analysis since they point to discussions already analyzed. Secondly, we used the Stack Exchange Data Explorer \footnote{https://data.stackexchange.com/}, which is a website where it is possible to run queries against a weekly updated version of the SE databases. We executed our queries in the database version of April 14, 2024. Through these queries, we found discussions that pointed to or were pointed to by the discussions at the start.

The BSB consisted of using the extracted links list to identify new discussions to include. The first step was to go through the links list and exclude discussions that do not fulfill the basic criteria such as eliminating incomplete and untrustworthy discussions. The next step was to remove discussions from the list that have already been examined in our previous studies~\cite{blind-1, blind-2, blind-3} either as a false positive or not. As a result, we were able to find 34 new candidate discussions.

The FSB consisted of identifying new discussions based on those with at least a link to a discussion within the start set. Next, we applied the same filters used in the BSB. This step rendered 50 new candidate discussions.

Regarding the related links, they are not available in the Stack Exchange Data Explorer or in the data dump, but they can be obtained through the Stack Exchange API \footnote{https://api.stackexchange.com/}. So, we created scripts to extract these links using the API. Similarly to the linked links extraction, we also found discussions that pointed to or were pointed to by the discussions in the start set by related links.

When applying the related snowballing, we applied similar steps to the one applied in the linked snowballing. As a result, we found 687 and 1,132 new candidate discussions for, respectively, BSB and FSB.

A large number of discussions as a result of the related snowballing was expected since the related links are built automatically by the SE platform. The resulting amount of discussions was considered unfeasible to be analyzed in this study. So, we performed one more filtering to reduce the amount of discussions. We extracted from the start set (the originally selected discussions) the average amount of answers and the average score per question. They were, respectively, four answers and a score eight per discussion. Thus, we filtered the related discussions where the number of answers was bigger or equal to four and the score of the question was bigger or equal to eight. After this final filtering, the number of resulting discussions for both BSB and FSB were, respectively, 104 and 156.


At the last step before the analysis, we grouped all the discussions found in two types of snowballing. During this process, we have counted how many discussions appeared in each combination of different snowballings, the numbers are presented in Table~\ref{tab:sb-combinations}. As expected the duplications between related BSB and related FSB was the most combination found (37 out of 47 duplications), since related links are built based in the similarities between questions. These duplications indicate that similar discussions point to each other via related links.

\begin{table}[]
\caption{Amount of duplicated discussions among snowballings.}
\begin{tabular}{|c|c|}
\hline
\textbf{SB combination}                                                                       & \textbf{\# of duplicated discussions} \\ \hline
Related BSB + Related FSB                                                                     & 37                         \\ \hline
\begin{tabular}[c]{@{}c@{}}Linked BSB + Related BSB +\\ Related FSB\end{tabular}              & 4                          \\ \hline
Linked BSB + Related FSB                                                                      & 2                          \\ \hline
\begin{tabular}[c]{@{}c@{}}Linked BSB + Linked FSB +\\ Related BSB + Related FSB\end{tabular} & 1                          \\ \hline
Linked BSB + Related BSB                                                                      & 1                          \\ \hline
Linked BSB + Linked FSB                                                                       & 1                          \\ \hline
Linked FSB + Related FSB                                                                      & 1                          \\ \hline
\end{tabular}
\label{tab:sb-combinations}
\end{table}

After grouping the duplicated discussions, we analyzed them first to avoid duplication of our analysis efforts. The final number of unique discussions for linked BSB, linked FSB, related BSB and related FSB were, respectively, 25, 47, 61, and 111. Including the 47 duplicated discussions, we end up with 291 unique new discussions that were analyzed qualitatively following the same approach discussed in Section~\ref{sec:start-set-and-analysis-process}.

In addition to the processes described in this section, we also counted the number of links for each one of the 291 discussions during related BSB and linked BSB. For example, the discussions with IDs 26070 and 28023 have linked links to 26011, and discussion 29724 has a related link pointing to discussion 26011, so 26011 has 3 citations. In the WL, the amount of citations of a paper is one of the metrics used to define the relevance of a paper. So, through this counting, we hope to find clusters of links that can indicate more relevant discussions within the data set.

Due to space limitations in this paper and our focus on the snowballing approach, we do not discuss the results of this qualitative analysis in detail. Furthermore, because of the large amount of data that needed to be analyzed, we decided to conduct only one SB iteration. The next section discusses the results of this snowballing iteration focusing on the validity of the discussions identified by each type of snowballing.

\section{Results and Discussion} \label{sec:result-and-discussion}

This section presents the main results obtained from the discussions analysis. These results are used to answer the proposed research questions. Our replication package is available at \url{https://zenodo.org/records/11123633}. There, the interested reader can find our complete data set.

\subsection{RQ1: How many new discussions does the snowballing find in relation to the original data set?}

After the removal of false positives during the qualitative analysis phase, our final sample of TD-related discussions on SEPM was reduced from 291 to 130 valid discussions. In relation to our previous studies that originated the start set, in which we had 108 valid discussions, it was an increase of 120\% of the data set.

None of the 291 discussions discovered through the snowballing had the terms 'debt' or 'shortcut' in their posts (questions and answers) and comments, nevertheless 130 of them were deemed to be valid. This result shows that the snowballing approach can be used to find new discussions effectively and expand the original data set.

Moreover, this also means that through the SB approach, we can bring to light new terms that can be considered in future studies. For example, when searching for the term 'motivate' among the valid discussions, we found 19 discussions containing the term, which is 15\% of the new data set. By our observation, the term 'motivate' seemed to be associated with team management issues. Thus, by using this term we may find discussions related to process debt. To make more clear our conclusion regarding the term 'motivate', the following are some examples of the titles of valid discussion that included the term:

\begin{itemize}
    \item ``How to deal with a team member who keeps missing deadlines?''
    \item ``How to get burned out team back engaged again?''
    \item ``Advice for dealing with a cowboy programmer in an agile team.''
    \item ``Dealing with a coworker who keeps making the same mistakes over and over.''
    \item ``Agile team missing commitments regularly and complaining about no trust.''
    \item ``How to motivate offshore teams and trust them to deliver?''
    \item ``Help - Technical team does not want to work in agile way.''
\end{itemize}

Regarding the most cited discussions during BSB, the IDs of the top most cited discussions (alongside the number of citations per type of link) were: \textbf{8286} (6 related), \textbf{11144} (6 related), \textbf{718} (4 related), \textbf{15505} (3 related) and \textbf{16372} (2 linked and 1 related). As we can see, the cluster links were mostly compound per related links, all discussions found via linked BSB had only one linked pointing to it, except for 16372 which had two links. Analyzing the most cited discussions, we can see that they have lots of upvotes, for example, discussion 16372 has alone 91 upvotes scattered between its posts. This means that they were deemed highly useful by the community, being considered more relevant. We can make a parallel between the relevance of a discussion and the relevance of a paper, where in both cases the amount of citations is the key metric for relevance. Finally, all these top-cited discussions were related to process debt and agile, which is a reflection of our data sets that have most discussions related to process debt and agile. These results show that, in a similar way to the WL, highly cited discussions tend to be more relevant.

\vspace{0.4cm}
\noindent\fbox{
\begin{minipage}{0.455\textwidth}
    \textbf{Finding \#1}: The SB conducted rendered 20\% more valid discussion than the original study, and hence it provides an added value to the original study. Besides that, we also have evidence that more cited discussions tend to be more relevant.
\end{minipage}
} 

\subsection{RQ2: How precise are snowballing techniques in finding new valid discussions?}

When conducting SB in SLRs, precision is a key metric to evaluate its efficiency. Precision refers to the number of included papers concerning the total number of candidate papers examined \cite{wohlin2014guidelines}. A high precision ensures that the gathered literature remains closely aligned with the topic of interest. This metric is essential for maintaining focus and validity in SLRs. To measure the precision of our study, we used a similar approach to the SLRs, considering discussions as papers and valid discussions as included papers.

As previously discussed, we detected 130 valid discussions in the 291 new discussions found during the snowballings, this was a precision of 45\% among the snowballings. The database search that originated the start set had a precision of 48\% (108 valid out of 226 discussions). The snowballing had a very similar precision to the database search, only a 3\% difference, especially when considering that the SB data set was 29\% larger. The combined precision of the database search and the SB approach is (130 + 108) / (291 + 226) = 46\%.

The detailed numbers for each type of SB, including the duplications in the accounting, are:
\begin{itemize}
    \item \textbf{Linked BSB:} 34 candidates discussions and 15 valid, i.e. precision = 15 / 34 = 44\%.
    \item \textbf{Linked FSB:} 50 candidates discussions and 19 valid, i.e. precision = 19 / 50 = 38\%.
    \item \textbf{Related BSB:} 104 candidates discussions and 59 valid, i.e. precision = 59 / 104 = 56\%.
    \item \textbf{Related FSB:} 156 candidates discussions and 69 valid, i.e. precision = 69 / 156 = 44\%.
\end{itemize}

The linked SB had an overall precision of (15 + 19) / (34 + 50) = 40\%. On the other hand, the related SB had an overall precision of (59 + 69) / (104 + 156) = 49\%. The related SB showed to be 9\% more precise than the linked SB even with three times more discussions. It appears that discussions found via related links tend to be more effective because related links between discussions are built based on the similarity to the \underline{valid discussions} from the start set.

\begin{figure}[htp]
\centerline{\includegraphics[width=\linewidth]{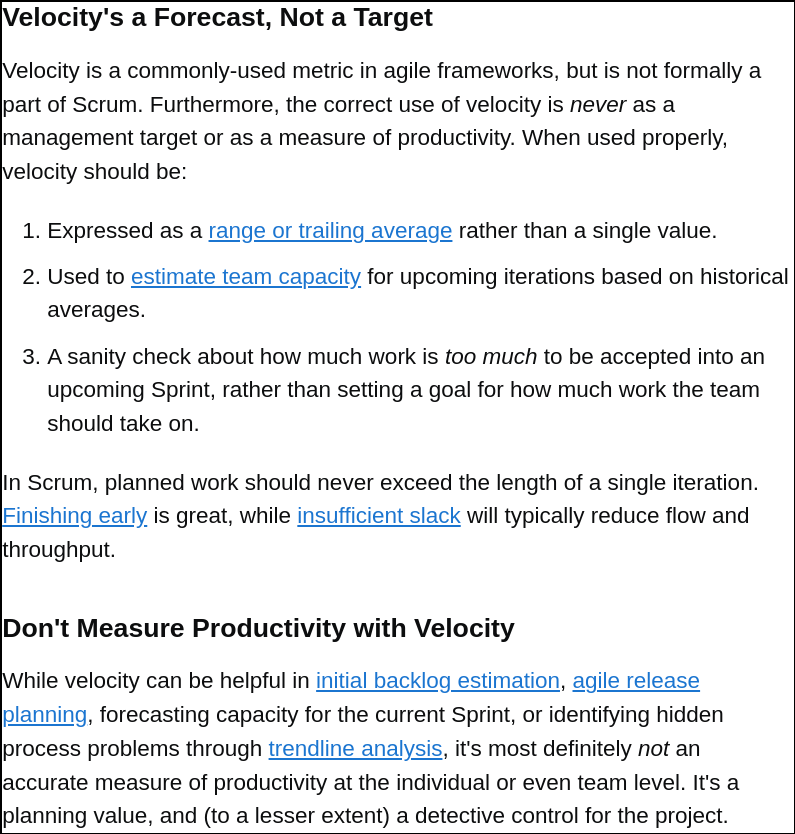}}
\caption{Example of linked links usage.}
\label{fig-linked-links}
\end{figure}

It is also worth mentioning the way linked links were used by practitioners. During our qualitative analysis, we detected cases where the SEPM users used these links to point to more theoretical discussions to reinforce their opinions or to link to more detailed discussions about a topic. In Figure~\ref{fig-linked-links}, we have an excerpt of a discussion answer from our start set where the user used 8 links (highlighted in blue), where 2 of these links are for blog posts and 6 of them are linked links, and these links are related to various topics such as sprint delivery planning and team velocity management. The links in the scenario shown in Figure~\ref{fig-linked-links} are used to indicate more in-depth materials about the topics, meanwhile, the answer itself is focused in a broader scope with a focus on providing a solution to the TD instance presented in the question. Regarding the precision of the 6 discussions discovered via the linked links presented in Figure~\ref{fig-linked-links}, only 2 of them were considered valid (33 \% precision). The lower precision of the linked links compared to the related links in the SB can also be attributed to the context and rules of the analysis. Despite being considered false positives, theoretical discussions also have importance, since they show the perspective of practitioners concerning some topics. 

Observe that, in a certain way, link-based SB is similar to our use of references in white literature review SB. When we are snowballing over a paper's references, we discard many because they discuss issues not central to the SLR inclusion criteria. Relation-based SB, on the other hand, selects content by global similarity. This process returns more material (better recall) and with a similar precision. From this perspective, our results indicate that researchers interested in systematic literature reviews would also benefit from similarity-based search mechanisms in digital libraries of white literature. 

\vspace{0.4cm}
\noindent\fbox{
\begin{minipage}{0.455\textwidth}
    \textbf{Finding \#2}: Related SB found more discussions than linked SB, but they are complementary. It is concluded that the full-fledged hybrid search strategy is the best, although it comes at the cost of more work.
\end{minipage}
} 


\section{Threats to Validity} \label{sec:threats-to-validity}

Below we present the threats to the validity of our research, grouping them according to the concepts specified by Wohlin \textit{et al.}~\cite{wohlin2012experimentation}.

\textbf{Construct:} One threat is selecting and analyzing the discussions coming from linked links and related links directly from SEPM. This risk was mitigated by using both the discussion evaluation process and the set of discussions from our previous studies~\cite{blind-1,blind-2,blind-3} as the basis for this work. Another threat can be attributed to how the SB procedure was conducted. We mitigate this by following the procedures defined by Wohlin \textit{et al.}~\cite{wohlin2014guidelines}, adapting some concepts to our context when necessary.

\textbf{Internal:} The process used for analyzing qualitatively the discussions can represent a threat in our study. To reduce this thread, the procedures portrayed in Figure~\ref{fig-analysis-process} were performed by two researchers individually. Besides, a third researcher was inserted to resolve the divergences identified in the consensus phase.

\textbf{External:} Regarding generalizing the conclusions, the study was based on a representative sample of SEPM, a well-known Q\&A platform focused on project management discussions, where practitioners discuss day-to-day issues. Although we used SEPM data, we cannot guarantee that all users are practitioners. This is mitigated by the fact that SEPM discussions are contextualized in the software management area. Another threat arises from the risk that the newly selected discussions via SB are unrelated to TD. To mitigate this threat, we qualitatively analyzed the discussions using the same approach from our previous studies~\cite{blind-1,blind-2,blind-3} related to analyzing TD on SEPM. An even more critical threat stems from the fact that SEPM does not represent all software engineeering Q\&A sites, albeit the fact that it belongs to Stack Exchange, which hosts a large family of software-related Q\&A sites. We argue that the resources we used from this platform (direct links and similarity searches) will eventually be available in several Q\&A platforms.

\textbf{Conclusion:} Lastly, there is a risk that, even if we applied the same approach to the analysis and interpretation of the discussions, there is the possibility of producing different results. This risk stems from the subjectivity inherent in the process. We mitigated it by the consensus procedures used during the analysis process presented in Section~\ref{sec:research-method}. Nonetheless, the similarity search algorithm used by the Stack Exchange Platform can evolve and affect the results we have obtained in this study, hopefully providing even better precision and recall of the selected discussions. 

\section{Conclusion} \label{sec:conclusion}

This work investigates how snowballing can be conducted on Stack Exchange websites. To the best of our knowledge, this is the first study that defines and analyzes the use of SB in discussion forums. We also identified two types of links between discussions, linked and related. By performing mining and qualitative analysis of these links in the SEPM forum, we provide a general approach that can be used in any of the SE websites, such as Stack Overflow.

This work also shows that SB can be effective and precise in finding new valid discussions. These results indicate that the use of SB can complement the use of text-based searches on the SE websites. In addition to new discussions, SB by similarity (related-based) can also bring to light search terms that can be used to expand the initial data set.

Regarding the number of citations of discussions, we also found evidence that often cited discussions tend to be more relevant inside the community. This is aligned with the WL, where highly cited papers are also considered relevant for the research community.

Finally, we share our insights regarding each type of discussion link. We discussed that linked links are used beyond linking similar discussions, they are also used to indicate more theoretical discussions to corroborate the presented opinion or affirmation. This is very similar to how citations are used in the white literature. Analyzing these types of discussions can provide the point of view of the community about some topic. On the other hand, we discussed that related links are more precise and have better recall when searching for valid discussions, since they point toward discussions similar to the ones at the start set. Finally, we commented that despite the differences between them, both types of links should be considered since they drew new valid discussions.

For researchers, our new method can support research efforts in GL. The proposed SB approach can be used to expand any study about discussion analysis on the SE websites. For example, we can use SB to expand the study of Gama \textit{et al.}~\cite{Gama2020} to enlarge their findings regarding how developers identify TD items on Stack Overflow. Lastly, our findings regarding linked SB and related SB can also motivate new research. For example, a comparison between the SE similarity algorithm with other algorithms, or the use of similarity search for SB in systematic literature reviews. 

In future work, we intend to (1) perform more SB iterations to expand our study, and (2) investigate the use of other similarity algorithms in comparison to the SE in the analysis process.


\bibliographystyle{ACM-Reference-Format}
\bibliography{references}

\appendix

\end{document}